\documentclass[12pt]{iopart}
\usepackage{graphicx}
\usepackage{latexsym}

\usepackage{iopams}

\def\h0units{\mathrm{km\,s^{-1}\,Mpc^{-1}}}
\def\sun{\hbox{$\odot$}}
\def\araa{ARA\&A  }

\def\mnras{MNRAS\,  }

\begin{document}
\title
{
Classical and relativistic
flux of  energy conservation  in astrophysical jets
}
\vskip  1cm
\author     {Lorenzo Zaninetti}
\address    {Physics Department  ,
 via P.Giuria 1,\\ I-10125 Turin,Italy }
\ead {zaninetti@ph.unito.it}

\begin {abstract}
The conservation of the energy flux  in turbulent jets 
which propagate in the intergalactic medium (IGM) 
allows deducing the  law of motion 
in the classical and relativistic cases.
Three types of IGM are considered: constant density, 
hyperbolic and inverse power law decrease of density.
An analytical law for the evolution of the magnetic field 
along the radio-jets is deduced using a linear relation 
between the magnetic pressure and the rest density.
Astrophysical applications are made to the centerline
intensity of synchrotron emission in NGC315 and to 
the magnetic field of 3C273.
\end{abstract} 
\vspace{2pc}

\maketitle

\section{Introduction}
The analysis  of turbulent jets in the laboratory
offers the possibility of applying the theory  of turbulence
to some well defined  experiments, see
\cite{Reynolds1883,Reynolds1894}.
The experiments of Reynolds  can be seen
in \cite{vanDyke1982}.
Analytical results for the theory  of turbulent jets
can be found  in \cite{goldstein,landau,pope2000,foot}.
Recently the analogy between laboratory
jets and extragalactic radio-jets has been pointed out, 
see
\cite{Lebedev2011,Suzuki-Vidal2012}.
We briefly recall that the theory of 
`round turbulent jets' can be defined in terms 
of the velocity at the nozzle, the diameter of the nozzle, 
and the viscosity, see Section 5 in \cite{pope2000};
as an example the gradients in pressure are not considered. 
The application of the theory of turbulence to extragalactic 
radio-jets produces a great number of questions to be solved 
because we do not observe the turbulent phenomena 
but the radio features which have properties 
similar to the laboratory's turbulent jets, i.e. similar opening angles.
We now  pose
the following questions.
\begin{itemize}
\item Is it possible to apply the conservation of the flux 
     of energy in order to derive the equation of 
      motion for radio-jets in the cases of constant 
     and variable density of 
     the surrounding medium?
\item Can we extend the conservation of the flux of energy 
    to the relativistic regime?
\item Can we model the behaviour of the magnetic field 
      and the intensity of synchrotron emission as functions 
      of the distance from the parent nucleus?
\item Can we model the back reaction on the equation of motion 
      for turbulent jets due to radiative losses?
\end{itemize}      
In order to answer these questions,
we derive the  differential equations
which model the classical and relativistic  
conservation of the energy flux  
for a  turbulent jet in  the presence of different  
types of medium, see Sections
\ref{secclassic} and
\ref{secrelativistic}.
Section \ref{seclosses} presents 
classical and relativistic parametrizations 
of the radiative losses
as well as the evolution of the magnetic field.

\section{Energy conservation}
\label{secclassic}

The conservation of the energy    flux in a turbulent jet
requires the perpendicular section to the motion along the
Cartesian $x$-axis, $A$
\begin {equation}
A(r)=\pi~r^2
\end{equation}
where $r$ is the radius of the jet.
The
section  $A$ at  position $x_0$  is
\begin {equation}
A(x_0)=\pi ( x_0   \tan ( \frac{\alpha}{2}))^2
\end{equation}
where   $\alpha$  is the opening angle and
$x_0$ is the initial position on the $x$-axis.
At position $x$ we have
\begin {equation}
A(x)=\pi ( x   \tan ( \frac{\alpha}{2}))^2
\quad .
\end{equation}
The conservation  of energy flux states that
\begin{equation}
\frac{1}{2} \rho(x_0)  v_0^3   A(x_0)  =
\frac{1}{2} \rho(x  )   v(x)^3 A(x)
\label{conservazioneenergy}
\end {equation}
where $v(x)$ is the velocity at  position $x$ and
$v_0(x_0)$   is the velocity at  position $x_0$,
see Formula A28 in \cite{deyoung}.

The selected physical units are
pc for length  and  yr for time;
with these units, the initial velocity $v_{{0}}$
is  expressed in pc yr$^{-1}$,
 1 yr = 365.25 days.
When the initial velocity is expressed in
km\,s$^{-1}$, the multiplicative factor $1.02\times10^{-6}$
should be applied in order to have the velocity expressed in
pc yr$^{-1}$.

\subsection{Constant  density}

\label{classicalconstant}

In the case of constant density of the intergalactic medium (IGM)
along the $x$-direction,
the  law of conservation of the
energy flux, as given by Eq. (\ref{conservazioneenergy}),
can be written as a  differential equation
\begin{equation}
\left( {\frac {\rm d}{{\rm d}t}}x \left( t \right)  \right) ^{3}
 \left( x \left( t \right)  \right) ^{2}-{v_{{0}}}^{3}{x_{{0}}}^{2}
=0
\quad .
\label{diffequationclassic}
\end{equation}
The analytical  solution of the previous differential
equation can be found by imposing $x=x_0$ at t=0,
\begin{equation}
x(t) =
\frac{1}{3}\,{3}^{2/5}\sqrt [5]{{x_{{0}}}^{2} \left( 5\,tv_{{0}}+3\,x_{{0}}
 \right) ^{3}}
\quad .
\label{energyxtconstant}
\end{equation}
The asymptotic approximation
is
\begin{equation}
x(t) \sim
\frac{1}{3}\,{3}^{2/5}{5}^{3/5}\sqrt [5]{{v_{{0}}}^{3}{x_{{0}}}^{2}}{t}^{3/5}
\quad .
\end{equation}
The velocity is
\begin{equation}
v(t) =
{\frac {{3}^{2/5}{x_{{0}}}^{2} \left( 5\,tv_{{0}}+3\,x_{{0}} \right) ^
{2}v_{{0}}}{ \left( {x_{{0}}}^{2} \left( 5\,tv_{{0}}+3\,x_{{0}}
 \right) ^{3} \right) ^{4/5}}}
\label{energyvtconstant}
\end{equation}
and its asymptotic approximation
\begin{equation}
v(t) \sim
\frac{1}{5}\,{\frac {{3}^{2/5}\sqrt [5]{125}{x_{{0}}}^{2}{v_{{0}}}^{3} \left(
{t}^{-1} \right) ^{2/5}}{ \left( {v_{{0}}}^{3}{x_{{0}}}^{2} \right) ^{
4/5}}}
\quad  .
\end{equation}
The velocity as a function of the distance is
\begin{equation}
v(x) = {\frac {{x_{{0}}}^{2/3}v_{{0}}}{{x}^{2/3}}}
\quad .
\label{energyvx}
\end{equation}

A first comparison can be made  with the laboratory
data on turbulent jets of  \cite{Mistry2014}
where the  velocity of the turbulent jet
at the nozzle diameter, $D_j$=1,
is $v_0 =2.53$\ m s$^{-1}$
and at  $D_j$=50 the  centerline  velocity
is $v =0.314$\ m s$^{-1}$.
The formula (\ref{energyvx})  with $x_0=1$ and $x=50$
gives  an averaged velocity of $v=0.186$\ m s$^{-1}$
which multiplied by 2 gives $v =0.372$\ m s$^{-1}$.
This multiplication by 2 has been done because
the turbulent jet  develops a profile of velocity in the direction
perpendicular  to the jet's main axis and therefore
the centerline velocity is  approximately double that of the
averaged velocity.
The transit time, $t_{tr}$, necessary to travel a distance
of $x_{max}$ can be derived from Eq. (\ref{energyxtconstant})
\begin{equation}
t_{tr} =
\frac
{
3\,\sqrt [3]{{x_{{\max}}}^{2}x_{{0}}}x_{{\max}}-3\,{x_{{0}}}^{2}
}
{
5\,x_{{0}}v_{{0}}
}
\quad .
\label{energytransit}
\end{equation}

An astrophysical  test can be  performed on a typical distance
of 15 kpc relative  to the jets in 3C\,31,
see Figure 2 in \cite{laing2002}.
On inserting  $x=15000\,$pc$=15$\ kpc, $x_0=100$\ pc,
and $v_0=10000$\ km s$^{-1}$ we obtain a transit time
of  $t_{tr}=2.488\,10^7$\ yr.

The rate of mass flow at the point $x$, $\dot {m}(x)$, is
\begin{equation}
\dot {m}(x) =
\rho   v(x)  \pi ( x   \tan ( \frac{\alpha}{2}))^2
\end{equation}
and the astrophysical version is
\begin{equation}
\dot {m}(x) =
0.0237n {x}^{4/3} \left( \tan
 \left( \alpha/2 \right)  \right) ^{2}{x_{{0}}}^{2/3}\beta_{{0}}\,\frac{\it
M_{\sun}}{yr}
\end{equation}
where $x$ and $x_0$ are expressed in pc,
$n $ is the number density of protons   expressed  in
particles~cm$^{-3}$,
$M_{\sun}$ is the solar mass
and $\beta_0=\frac{v_0}{c}$.
The previous formula indicates  that
the rate of transfer of particles  
is not constant along the
jet but increases  $\propto x^{4/3}$.

\subsection{An hyperbolic  profile of the density}

Now the density  is  assumed to decrease as
\begin{equation}
\rho = \rho_0  (\frac{x_0}{x})
\label{profhyperbolic}
\end{equation}
where  $\rho_0=0$ is the density at  $x=x_0$.
The differential equation that models the
energy  flux is
\begin{equation}
x_{{0}}x \left( t \right)  \left( {\frac {\rm d}{{\rm d}t}}x \left( t
 \right)  \right) ^{3}-{v_{{0}}}^{3}{x_{{0}}}^{2}
= 0
\end{equation}
and its analytical solution
is
\begin{equation}
x(t) =
\frac{1}{3}\,\sqrt [4]{3}\sqrt [4]{x_{{0}} \left( 4\,tv_{{0}}+3\,x_{{0}}
 \right) ^{3}}
\quad .
\label{xthyperbolic}
\end{equation}
The asymptotic  approximation
is
\begin{equation}
x(t)  \sim
\frac{2}{3} \,\sqrt [4]{3}\sqrt {2}\sqrt [4]{{v_{{0}}}^{3}x_{{0}}}{t}^{3/4}
\quad .
\label{xthyperbolicasympt}
\end{equation}
The analytical solution for the velocity
is
\begin{equation}
v(t) =
{\frac {\sqrt [4]{3}x_{{0}} \left( 4\,tv_{{0}}+3\,x_{{0}} \right) ^{2}
v_{{0}}}{ \left( x_{{0}} \left( 4\,tv_{{0}}+3\,x_{{0}} \right) ^{3}
 \right) ^{3/4}}}
\label{vthyperbolic}
\end{equation}
and its asymptotic  approximation is
\begin{equation}
v(t) \sim
\frac{1}{4}\,{\frac {\sqrt [4]{3}\sqrt [4]{64}x_{{0}}{v_{{0}}}^{3}\sqrt [4]{{t
}^{-1}}}{ \left( {v_{{0}}}^{3}x_{{0}} \right) ^{3/4}}}
\quad  .
\end{equation}

The transit time can be  derived from
Eq. (\ref{xthyperbolic})
\begin{equation}
t_{tr} =
\frac
{
3\,\sqrt [3]{x_{{\max}}{x_{{0}}}^{2}}x_{{\max}}-3\,{x_{{0}}}^{2}
}
{
4\,x_{{0}}v_{{0}}
}
\end{equation}
and with $x=15000$\ pc$=15$\ kpc, $x_0=100$\ pc,
and $v_0=10000$\ km s$^{-1}$  as in Section
\ref{classicalconstant},
we have  $t_{tr} =5.848\,10^6 $\ yr.

\subsection{An inverse power law  profile of the density}

\begin{figure*}
\begin{center}
\includegraphics[width=7cm]{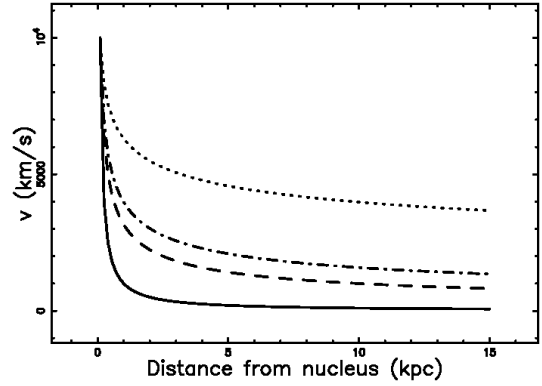}
\end {center}
\caption
{
Classical velocity   as a function
of  the distance from the nucleus  when
$x_0$ =100~pc and $v_0=10000\,$km s$^{-1}$:
$\delta =0$   (full line),
$\delta =1$   (dashes),
$\delta =1.2$ (dot-dash-dot-dash)
and
$\delta =1.6$ (dotted).
}
\label{energyveldelta}
    \end{figure*}

Here, the density  is  assumed to decrease as
\begin{equation}
\rho = \rho_0  (\frac{x_0}{x})^{\delta}
\label{profpower}
\end{equation}
where  $\rho_0$ is the density at  $x=x_0$.
The differential equation which  models the
energy  flux is
\begin{equation}
\frac{1}{2}\, \left( {\frac {x_{{0}}}{x}} \right) ^{\delta} \left( {\frac
{\rm d}{{\rm d}t}}x \left( t \right)  \right) ^{2}{x}^{2}
- \frac{1}{2}\,{v_{{0}
}}^{2}{x_{{0}}}^{2}
=0
\quad  .
\end{equation}
There is no analytical solution, and we simply express
the velocity as a function of the position, $x$,
\begin{equation}
v(x) =
{\frac {x_{{0}}v_{{0}}}{x}{\frac {1}{\sqrt { \left( {\frac {x_{{0}}}{x
}} \right) ^{\delta}}}}}
\label{velocitypower}
\end{equation}
see  Figure \ref{energyveldelta}

The rate of mass flow at the point $x$ is
\begin{equation}
\dot {m}(x) =
\rho_{{0}}\sqrt { \left( {\frac {x_{{0}}}{x}} \right) ^{\delta}}\pi \,
x \left( \tan \left( \alpha/2 \right)  \right) ^{2}x_{{0}}v_{{0}}
\label{mxpower}
\end{equation}
and the astrophysical version is
\begin{equation}
\dot {m}(x) =
 0.0237\,n_0\sqrt { \left(  1.0\,{\frac {x_{{0}}}{x}} \right) ^{
\delta}}x \left( \tan \left( \alpha/2 \right)  \right) ^{2}x_{{0}}
\beta_{{0}}  \frac{{\it M_{\sun}}}{yr}
\end{equation}
where
$n_0 $ is the number density of protons   expressed  in
particles~cm$^{-3}$ at $x_0$.
The previous formula indicates  that the
rate of transfer of particles   scales
$\propto x^{1-\frac{1}{2}\delta}$
and therefore at $\delta=2$ is constant.

\section{Relativistic turbulent jets}
\label{secrelativistic}

The conservation of the   energy flux in special relativity (SR)
in  the presence of a  velocity $v$ along one direction
states that
\begin{equation}
A(x) \frac { 1}{ 1 -\frac {v^2}{c^2}} (e_0 +p_0) v = cost
\label{enthalpy}
\end{equation}
where $A(x)$ is the considered area in the direction perpendicular
to the motion,
$c$ is the speed of light,
$e_0= c^2 \rho$ is the energy density in the rest 
frame of the moving fluid,
and $p_0$ is the pressure in the rest frame 
of the moving fluid,
see  formula A31 in
\cite{deyoung}.
In accordance with the current models of classical turbulent jets,
we  insert $p_0=0$  and
the   conservation law
for relativistic energy flux
is
\begin{equation}
\rho  c^2 v \frac { 1}{ 1 -\frac {v^2}{c^2} } A(x) = cost
\quad .
\label{relativisticflux}
\end{equation}
Our physical units are
pc for length  and  yr for time, and
in these units, the speed of light is
$c=0.306$  \ pc \ yr$^{-1}$.
A discussion of the mass--energy equivalence principle in fluids 
can be found in \cite{Palacios2015a}. 

\subsection{Constant  density in SR}

The  conservation of the relativistic energy flux
when the density is constant can be written
as a differential equation
\begin{eqnarray}
{\rho\,{c}^{2} \left( {\frac {\rm d}{{\rm d}t}}x \left( t \right)
 \right) \pi \, \left( x \left( t \right)  \right) ^{2} \left( \tan
 \left( \frac{\alpha}{2} \right)  \right) ^{2} \left( 1-{\frac { \left( {\frac
{\rm d}{{\rm d}t}}x \left( t \right)  \right) ^{2}}{{c}^{2}}}
 \right) ^{-1}}
 \nonumber \\
 -{\rho\,{c}^{2}v_{{0}}\pi \,{x_{{0}}}^{2} \left( \tan
 \left( \frac{\alpha}{2} \right)  \right) ^{2} \left(1 -{\frac {{v_{{0}}}^{2}}{{c
}^{2}}} \right) ^{-1}}=0
\label{eqndiffrel}
\quad  .
\end{eqnarray}
An analytical solution of the previous differential equation
at the moment of writing does not
exist but  we can
provide a  power series solution of the form
\begin{equation}
x(t) = a_0 +a_1  t +a_2 t^2 +a_3  t^3 + \dots
\label{xtrelseries}
\end{equation}
see  \cite{Tenenbaum1963,Ince2012}.
The coefficients $a_n$ up to order 4  are
\begin{eqnarray}
a_0=&  x_{{0}}      \nonumber \\
a_1=&  v_{{0}}    \nonumber  \\
a_2=&  \frac{1}{3}\,{\frac {{v_{{0}}}^{3} \left( 5\,{c}^{6}-11\,{c}^{4}{v_{{0}}}^{2}+
3\,{c}^{2}{v_{{0}}}^{4}+3\,{v_{{0}}}^{6} \right) }{{x_{{0}}}^{2}
 \left( {c}^{2}+{v_{{0}}}^{2} \right)  \left( {c}^{4}+2\,{c}^{2}{v_{{0
}}}^{2}+{v_{{0}}}^{4} \right) }}
    \nonumber \\
a_3=&
\frac{1}{3}\,{\frac {{v_{{0}}}^{3} \left( 5\,{c}^{6}-11\,{c}^{4}{v_{{0}}}^{2}+
3\,{c}^{2}{v_{{0}}}^{4}+3\,{v_{{0}}}^{6} \right) }{{x_{{0}}}^{2}
 \left( {c}^{2}+{v_{{0}}}^{2} \right)  \left( {c}^{4}+2\,{c}^{2}{v_{{0
}}}^{2}+{v_{{0}}}^{4} \right) }}
\quad .
\end{eqnarray}

\begin{figure*}
\begin{center}
\includegraphics[width=7cm]{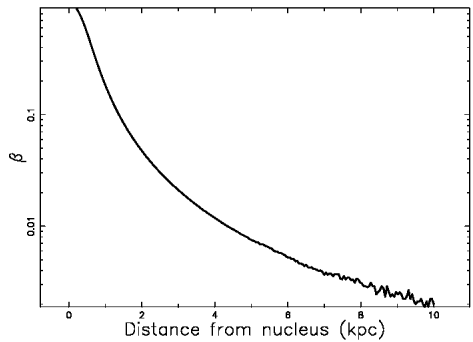}
\end {center}
\caption
{
Relativistic $\beta$   as a function
of  the distance from the nucleus  when
$x_0$ =200~pc and $\beta_0$ =0.9 in the case of constant
density.
}
\label{energybeta}
    \end{figure*}

In order to find  a numerical solution of the above
differential  equation
we isolate the velocity from
Eq.~(\ref{eqndiffrel})
\begin{eqnarray}
v(x;x_0,\beta_0,c) = \nonumber \\
\frac{1}{2}\,{\frac { \left( {\beta_{{0}}}^{2}{x}^{2}-{x}^{2}+\sqrt {{x}^{4}{
\beta_{{0}}}^{4}-2\,{x}^{4}{\beta_{{0}}}^{2}+4\,{\beta_{{0}}}^{2}{x_{{0
}}}^{4}+{x}^{4}} \right) c}{\beta_{{0}}{x_{{0}}}^{2}}}
\,
\label{vxrel}
\end{eqnarray}
where $\beta_0=\frac{v_0}{c}$
and  separate the variables
\begin{eqnarray}
\int_{x_0}^x
2\,{\frac {\beta_{{0}}{x_{{0}}}^{2}}{ \left( {\beta_{{0}}}^{2}{x}^{2}-
{x}^{2}+\sqrt {{x}^{4}{\beta_{{0}}}^{4}-2\,{x}^{4}{\beta_{{0}}}^{2}+4
\,{\beta_{{0}}}^{2}{x_{{0}}}^{4}+{x}^{4}} \right) c}}
dx  \nonumber \\= \int_0^t  dt
\quad .
\end{eqnarray}
The indefinite integral on the left side 
of the previous equation
has an analytical expression
\begin{equation}
I(x;\beta_0,c,x_0)=
\frac{AN}{AD}
\end{equation}
where
\begin{eqnarray}
AN =& \nonumber \\
2\,{\beta_{{0}}}^{3}{x_{{0}}}^{6}\sqrt {2}\sqrt {4-2\,{\frac {i\beta_{
{0}}{x}^{2}}{{x_{{0}}}^{2}}}+2\,{\frac {i{x}^{2}}{\beta_{{0}}{x_{{0}}}
^{2}}}}\sqrt {4+2\,{\frac {i\beta_{{0}}{x}^{2}}{{x_{{0}}}^{2}}}-2\,{
\frac {i{x}^{2}}{\beta_{{0}}{x_{{0}}}^{2}}}}\times
~& \nonumber \\
 \times {\it F} ( 1/2
\,x\sqrt {2}\sqrt {{\frac {i ( {\beta_{{0}}}^{2}-1 ) }{
\beta_{{0}}{x_{{0}}}^{2}}}},i )
\nonumber \\ -{\beta_{{0}}}^{3}{x_{{0}}}^{2}{
x}^{3}\sqrt {{\frac {i\beta_{{0}}}{{x_{{0}}}^{2}}}-{\frac {i}{\beta_{{0
}}{x_{{0}}}^{2}}}}\sqrt {{x}^{4}{\beta_{{0}}}^{4}-2\,{x}^{4}{\beta_{{0
}}}^{2}+4\,{\beta_{{0}}}^{2}{x_{{0}}}^{4}+{x}^{4}}
~&\nonumber \\
+\beta_{{0}}{x_{{0}}
}^{2}{x}^{3}\sqrt {{\frac {i\beta_{{0}}}{{x_{{0}}}^{2}}}-{\frac {i}{
\beta_{{0}}{x_{{0}}}^{2}}}}\sqrt {{x}^{4}{\beta_{{0}}}^{4}-2\,{x}^{4}{
\beta_{{0}}}^{2}+4\,{\beta_{{0}}}^{2}{x_{{0}}}^{4}+{x}^{4}}
&\nonumber \\
+{\beta_{{0
}}}^{5}{x_{{0}}}^{2}{x}^{5}\sqrt {{\frac {i\beta_{{0}}}{{x_{{0}}}^{2}}
}-{\frac {i}{\beta_{{0}}{x_{{0}}}^{2}}}}
-2\,{\beta_{{0}}}^{3}{x_{{0}}}
^{2}{x}^{5}\sqrt {{\frac {i\beta_{{0}}}{{x_{{0}}}^{2}}}-{\frac {i}{
\beta_{{0}}{x_{{0}}}^{2}}}}
~& \nonumber \\
+4\,{\beta_{{0}}}^{3}{x_{{0}}}^{6}x\sqrt {{
\frac {i\beta_{{0}}}{{x_{{0}}}^{2}}}-{\frac {i}{\beta_{{0}}{x_{{0}}}^{
2}}}}+\beta_{{0}}{x_{{0}}}^{2}{x}^{5}\sqrt {{\frac {i\beta_{{0}}}{{x_{
{0}}}^{2}}}-{\frac {i}{\beta_{{0}}{x_{{0}}}^{2}}}}
\end{eqnarray}
and
\begin{eqnarray}
AD= \nonumber \\
6\,c{\beta_{{0}}}^{2}{x_{{0}}}^{4}\sqrt {{\frac {i\beta_{{0}}}{{x_{{0}
}}^{2}}}-{\frac {i}{\beta_{{0}}{x_{{0}}}^{2}}}}\sqrt {{x}^{4}{\beta_{{0
}}}^{4}-2\,{x}^{4}{\beta_{{0}}}^{2}+4\,{\beta_{{0}}}^{2}{x_{{0}}}^{4}+
{x}^{4}}
\end{eqnarray}
where $i=\sqrt{-1}$
and
\begin{equation}
F(x;m)=\int_{0}^{x}\!{\frac {1}{\sqrt {1-{{\it t}}^{2}}\sqrt {1-{m}^{2
}{{\it t}}^{2}}}}\,{\rm d}{\it t}
\end{equation}
is the elliptic integral of the first kind,
see formula 17.2.7 in \cite{Abramowitz1965}.
Figure \ref{energybeta} shows the
behaviour of $\beta$ as  function
of the distance.

\begin{figure*}
\begin{center}
\includegraphics[width=7cm]{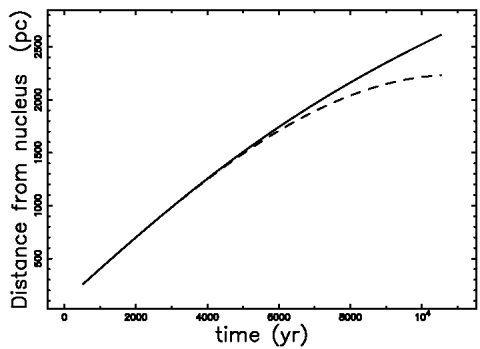}
\end {center}
\caption
{
Non-linear relativistic solution  as given
by Eq. (\ref{solrelativisticnl}) (full line)
and series solution
 as given
by Eq. (\ref{xtrelseries}) (dashed line)
when
$x_0$ =100 pc and $\beta_0$ =0.999.
}
\label{energyrelxt}
    \end{figure*}

A numerical  solution can be found  by solving
the following non-linear  equation
\begin{equation}
I(x;\beta_0,c,x_0)- I(x_0;\beta_0,c,x_0)=t
\label{solrelativisticnl}
\end{equation}
 and  Figure  \ref{energyrelxt}
presents a typical comparison with the series solution.
The relativistic rate of mass flow in the case of 
constant density is
\begin{eqnarray}
\dot {m}(x) =  & ~ \nonumber \\
\frac
{
\rho\, \left( {\beta_{{0}}}^{2}{x}^{2}-{x}^{2}+\sqrt {{x}^{4}{\beta_{{0
}}}^{4}-2\,{x}^{4}{\beta_{{0}}}^{2}+4\,{\beta_{{0}}}^{2}{x_{{0}}}^{4}+
{x}^{4}} \right) c\pi \,x \left( \tan \left( \alpha/2 \right)
 \right) ^{2}
}
{
\sqrt {2 \left( 1-{\beta_{{0}}}^{2} \right)  \left( {\beta_{{0}}}^{
2}{x}^{2}-{x}^{2}+\sqrt {{x}^{4}{\beta_{{0}}}^{4}-2\,{x}^{4}{\beta_{{0
}}}^{2}+4\,{\beta_{{0}}}^{2}{x_{{0}}}^{4}+{x}^{4}} \right) }
}
& ~
\end{eqnarray}

\subsection{Inverse power law  profile of density in SR }

The  conservation of the relativistic energy  flux
in the presence  of an inverse power law
density profile
as given by Eq. (\ref{profpower})
is
\begin{eqnarray}
{\rho_0\,{c}^{2} \left( {\frac {\rm d}{{\rm d}t}}x \left( t \right)
 \right) \pi \, \left( x \left( t \right)  \right) ^{2} \left( \tan
 \left( \frac{\alpha}{2} \right)  \right) ^{2} \left( {\frac {x_{{0}}}{x
 \left( t \right) }} \right) ^{\delta} \left( -{\frac { \left( {\frac
{\rm d}{{\rm d}t}}x \left( t \right)  \right) ^{2}}{{c}^{2}}}+1
 \right) ^{-1}} \nonumber \\
-{\rho_0\,{c}^{2}v_{{0}}\pi \,{x_{{0}}}^{2} \left( \tan
 \left( \frac{\alpha}{2} \right)  \right) ^{2} \left( -{\frac {{v_{{0}}}^{2}}{
{c}^{2}}}+1 \right) ^{-1}}=0
\quad .
\label{diffeqnrelpower}
\end{eqnarray}
This  differential equation does not have an analytical
solution. An expression for 
$\beta$ as a function of the distance
is
\begin{equation}
\beta(x) =
\frac{1}{2}\,{\frac {1}{\beta_{{0}}{x_{{0}}}^{2}} \left( {\beta_{{0}}}^{2}{x}^
{2} \left( {\frac {x_{{0}}}{x}} \right) ^{\delta}-{x}^{2} \left( {
\frac {x_{{0}}}{x}} \right) ^{\delta}+\sqrt {D} \right) }
\label{betadistance}
\end{equation}
with
\begin{eqnarray}
D= & \nonumber \\
\left(  \left( {\frac {x_{{0}}}{x}} \right) ^{\delta} \right) ^{2}{
\beta_{{0}}}^{4}{x}^{4}-2\, \left(  \left( {\frac {x_{{0}}}{x}}
 \right) ^{\delta} \right) ^{2}{\beta_{{0}}}^{2}{x}^{4}+ \left(
 \left( {\frac {x_{{0}}}{x}} \right) ^{\delta} \right) ^{2}{x}^{4}+4\,
{\beta_{{0}}}^{2}{x_{{0}}}^{4}
\quad .
\label{eqnd}
\end{eqnarray}

\begin{figure*}
\begin{center}
\includegraphics[width=7cm]{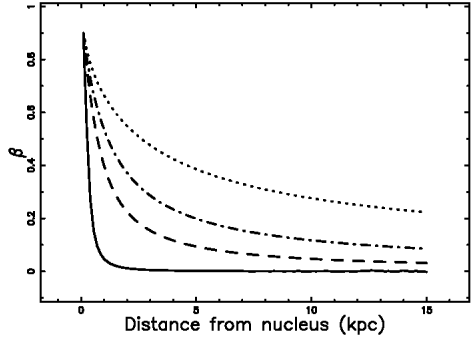}
\end {center}
\caption
{
Relativistic $\beta$ for the relativistic
energy flux conservation  as a function
of  the distance from the nucleus  when
$x_0$ =100~pc and $\beta_0$ =0.9:
$\delta =0$   (full line),
$\delta =1$   (dashes),
$\delta =1.2$ (dot-dash-dot-dash)
and
$\delta =1.4$ (dotted).
}
\label{betaxdeltaenergy}
    \end{figure*}

\begin{figure*}
\begin{center}
\includegraphics[width=7cm]{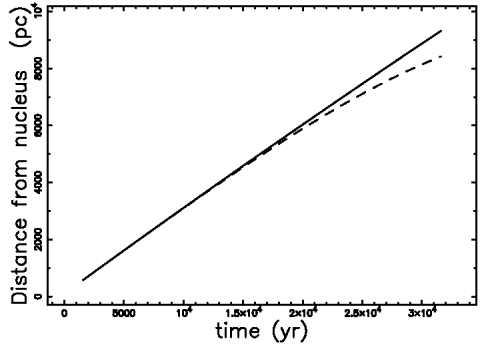}
\end {center}
\caption
{
Non-linear relativistic solution  as given
by Eq. (\ref{diffeqnrelpower}) (full line)
and series solution
 as given
by Eq. (\ref{xtseriesreldelta}) (dashed line)
when
$x_0$ =100 pc and $\beta_0$ =0.999.
}
\label{energyrelxtdelta}
    \end{figure*}

The behaviour of $\beta$ as a function of the distance
for different values of $\delta$ can be seen
in Figure \ref{betaxdeltaenergy}.
A  power series solution  for the above differential
equation (\ref{diffeqnrelpower})
up to order three  gives
\begin{eqnarray}
a_0=& x_{{0}}       \nonumber \\
a_1=& v_{{0}}      \nonumber  \\
a_2=& \frac{1}{2}\,{\frac {{v_{{0}}}^{2} \left( {c}^{2}\delta-\delta\,{v_{{0}}}^{2}-
2\,{c}^{2}+2\,{v_{{0}}}^{2} \right) }{x_{{0}} \left( {c}^{2}+{v_{{0}}}
^{2} \right) }}
\label{xtseriesreldelta}
\quad .
\end{eqnarray}
Figure  \ref{energyrelxtdelta}
shows a comparison between the numerical solution
of (\ref{diffeqnrelpower})
with the series solution.

The relativistic rate of mass flow in the case of an inverse power law for
the density is
\begin{equation}
\dot {m}(x) =
\frac
{
\rho_{{0}} \left( {\frac {x_{{0}}}{x}} \right) ^{\delta} \left( {\beta
_{{0}}}^{2}{x}^{2} \left( {\frac {x_{{0}}}{x}} \right) ^{\delta}-{x}^{
2} \left( {\frac {x_{{0}}}{x}} \right) ^{\delta}+\sqrt {D} \right) c
\pi \,{x}^{2} \left( \tan \left( \alpha/2 \right)  \right) ^{2}
}
{
2\,\beta_{{0}}{x_{{0}}}^{2}\sqrt {-1/4\,{\frac {1}{{\beta_{{0}}}^{2}{x
_{{0}}}^{4}} \left( {\beta_{{0}}}^{2}{x}^{2} \left( {\frac {x_{{0}}}{x
}} \right) ^{\delta}-{x}^{2} \left( {\frac {x_{{0}}}{x}} \right) ^{
\delta}+\sqrt {D} \right) ^{2}}+1}
}
\end{equation}
where  $\rho_0$ is the density at $x_0$ and
$D$ was defined in Eq.~(\ref{eqnd}).

\section{The losses}

\label{seclosses}
The previous analysis does not cover the radiative losses.
The astrophysical version
of the relativistic energy flux
as represented by   Eq.~(\ref{relativisticflux} ) is
\begin{equation}
\frac{dE}{dt} =
{ 1.348 \times
10^{49}}\,{\frac {n \beta_{{0}}{R_{{100}}}^{2
}}{1-{\beta_{{0}}}^{2}}}
\quad \frac {erg}{s}
\end{equation}
where $R_{100}$ is the radius of the jet expressed 
in units of 100 pc,
and $n$
is  the
number density of protons   expressed  in 
particles~cm$^{-3}$.
The above luminosity is 4--5 orders of magnitude 
too high for the radio sources here considered.
In order to explain this discrepancy, one model 
assumes that extragalactic  jets are much lighter 
than  the surroundings.
The second  model assumes that 
the observed intensity of radiation, $I_{\nu}$,   
at a given frequency $\nu$
is  a fraction of the energy flux  
\begin{equation}
I_{\nu}=\varepsilon \frac{dE}{dt}
\quad \frac {erg}{s}
\end{equation}
where $\varepsilon$ represents the efficiency of conversion
of the relativistic energy flux
into radiation.
At the moment of writing there is no 
exact evaluation
of  the efficiency of conversion.
We now outline two different models for the radiative losses
and a model for the magnetic field.

\subsection{Losses through recursion}

\begin{figure*}
\begin{center}
\includegraphics[width=7cm]{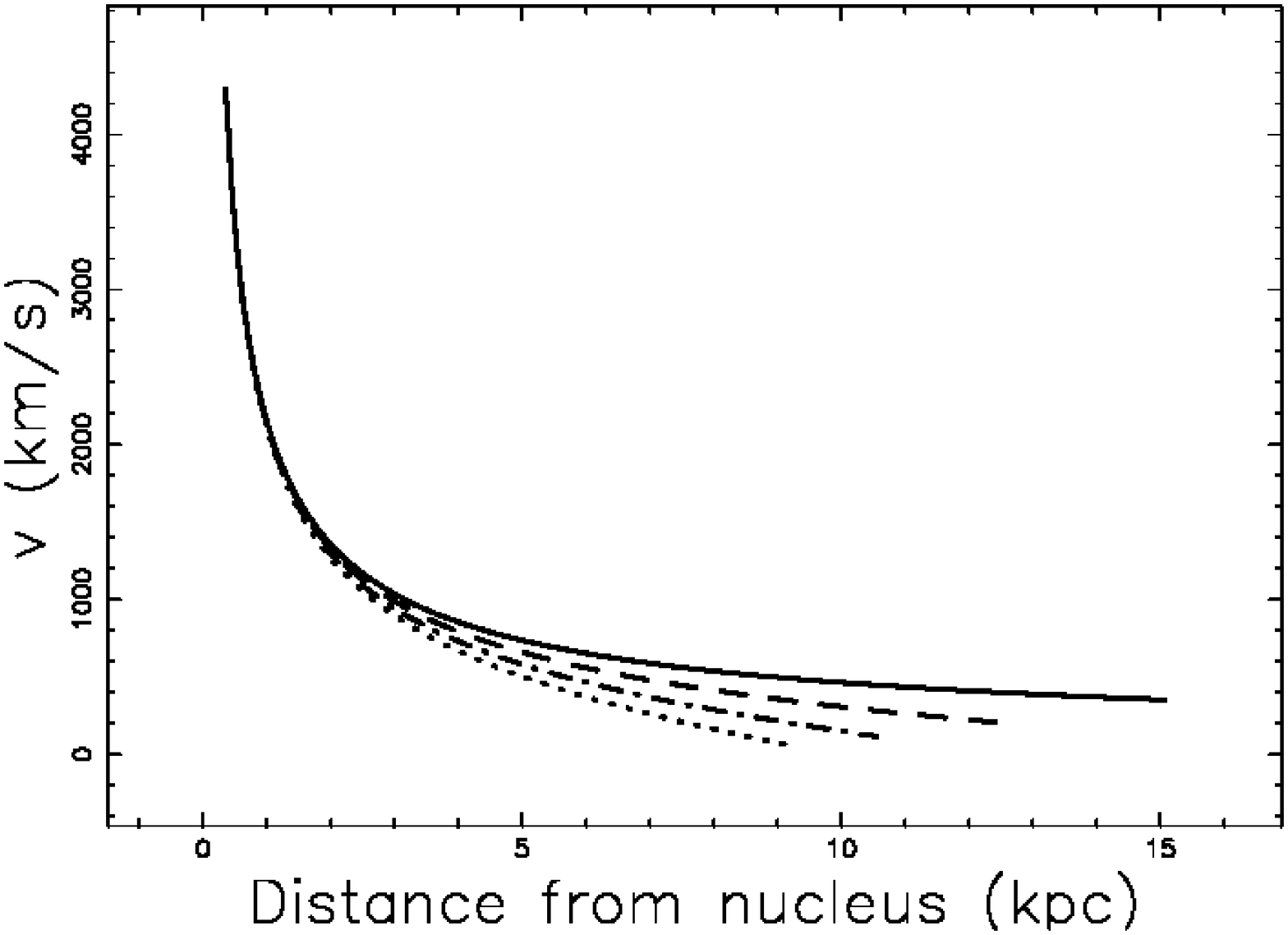}
\end {center}
\caption
{
Classical velocity   as a function
of  the distance from the nucleus  when
$x_0$ =100~pc,
$\Delta  t=2.5 10^4$\ yr
and $v_0=10000\,$km s$^{-1}$:
$\varepsilon =0$       (full line),
$\varepsilon =0.002$   (dashes),
$\varepsilon =0.004$   (dot-dash-dot-dash)
and
$\varepsilon =0.006$   (dotted).
}
\label{recursive_classic}
    \end{figure*}

In the classical case, with constant density,
we can model the radiative losses through the
following recursive equation
obtained by modifying  Eq.~(\ref{diffequationclassic})
\begin{equation}
\frac{1}{2} \rho   v_{n+1}^3
\pi (x_{n+1} \tan (\frac{\alpha}{2}))^2
+ \varepsilon \frac{1}{2} \rho   v_n^3   \pi (x_{n} \tan (\frac{\alpha}{2}))^2  =
\frac{1}{2} \rho   v_n^3   \pi (x_{n} \tan (\frac{\alpha}{2}))^2
\label{eqnrecursive}
\end{equation}
where
\begin{equation}
x_{n+1} = x_n + v_n \Delta t
\quad .
\end{equation}
Here $n$ starts from 0,
$v_n$ is the velocity  at the $n$th step,
$x_n$ is the position  at the $n$th step,
$\varepsilon$          is  the efficiency 
of conversion into radiation,
$\alpha$ is the jet's opening angle,
and
$\Delta \,t$ is the temporal step.
The velocity at step $n+1$ is
\begin{equation}
v_{n+1} = {\frac {{x_{{n}}}^{2/3}\sqrt [3]{1-\varepsilon}v_{{n}}}{ \left(
v_{{n}}
\Delta\,t+x_{{n}} \right) ^{2/3}}}
\quad .
\end{equation}
Figure \ref{recursive_classic}
shows the velocity as a function of the
distance; $\varepsilon \approx 10^{-4}$
does   not modify in an appreciable way  the velocity.

In the relativistic  case,
with constant density,
the radiative losses are modeled
by a   modification
of   Eq.~(\ref{eqndiffrel})
and the following recursive equation
for the velocity at step $n+1$  is obtained
\begin{equation}
v_{n+1}= \frac{N_n}{D_n}
\label{eqnrecursiverelativistic}
\end{equation}
where
\begin{eqnarray}
N_n = {c}^{4}{\Delta}^{2}{t}^{2}{v_{{n}}}^{2}-{c}^{2}{\Delta}^{2}{t}^{2}{v_{
{n}}}^{4}+2\,{c}^{4}\Delta\,tv_{{n}}x_{{n}}
\nonumber \\
-2\,{c}^{2}\Delta\,t{v_{{n}
}}^{3}x_{{n}}+{c}^{4}{x_{{n}}}^{2}-{c}^{2}{v_{{n}}}^{2}{x_{{n}}}^{2}-{
c}^{4}\sqrt {{\it S_n}}
\nonumber
\end{eqnarray}
\begin{eqnarray}
S_n = {\frac {{v_{{n}}}^{4} \left( v_{{n}}\Delta\,t+x_{{n}} \right) ^{4}}{{c
}^{4}}}+4\,{\frac {{v_{{n}}}^{2}{\varepsilon}^{2}{x_{{n}}}^{4}}{{c}^{2}}}
-8\,{\frac {{v_{{n}}}^{2}\varepsilon\,{x_{{n}}}^{4}}{{c}^{2}}}
\nonumber \\
-2\,{\frac
{{v_{{n}}}^{2} \left( v_{{n}}\Delta\,t+x_{{n}} \right) ^{4}}{{c}^{2}}}
+4\,{\frac {{v_{{n}}}^{2}{x_{{n}}}^{4}}{{c}^{2}}}+ \left( v_{{n}}
\Delta\,t+x_{{n}} \right) ^{4}
\nonumber
\end{eqnarray}
\begin{equation}
D_n =
2\,{c}^{2}v_{{n}}{x_{{n}}}^{2}\varepsilon-2\,{c}^{2}v_{{n}}{x_{{n}}}^{2}
\quad .
\nonumber
\end{equation}

\begin{figure*}
\begin{center}
\includegraphics[width=7cm]{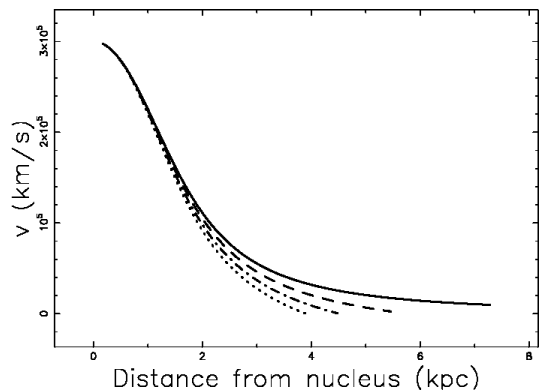}
\end {center}
\caption
{
Relativistic velocity   as a function
of  the distance from the nucleus  when
$x_0$ =100~pc,
$\Delta  t=250$\ yr,
and
$\beta_0= 0.999$:
$\varepsilon =0$       (full line),
$\varepsilon =0.002$   (dashes),
$\varepsilon =0.004$   (dot-dash-dot-dash)
and
$\varepsilon =0.006$   (dotted).
}
\label{recursive_rel}
    \end{figure*}

Figure \ref{recursive_rel} shows the relativistic
velocity as a function of the
distance  and  $\varepsilon$.

\subsection{The parametrization of the losses}

The radiative losses can also be modeled
by an `ad hoc' law for the available flux of kinetic energy,
which is  assumed  to decrease with  an inverse
power law of the type
$\propto \, (\frac{x_0}{x})^\eta$.
The resulting differential equation
in SR with constant density is
\begin{eqnarray}
{\rho\,{c}^{2} \left( {\frac {\rm d}{{\rm d}t}}x \left( t \right)
 \right) \pi \, \left( x \left( t \right)  \right) ^{2} \left( \tan
 \left( \frac{\alpha}{2} \right)  \right) ^{2} \left( 1-{\frac { \left( {\frac
{\rm d}{{\rm d}t}}x \left( t \right)  \right) ^{2}}{{c}^{2}}}
 \right) ^{-1}}
 \nonumber \\
 -{\rho\,{c}^{2}v_{{0}}\pi \,{x_{{0}}}^{2} \left( \tan
 \left( \frac{\alpha}{2} \right)  \right) ^{2} \left(1 -{\frac {{v_{{0}}}^{2}}{{c
}^{2}}} \right) ^{-1}} \bigl (\frac{x_0}{x}\bigr )^{\eta} =0
\label{eqndiffrellosses}
\quad  .
\end{eqnarray}

\begin{figure*}
\begin{center}
\includegraphics[width=7cm]{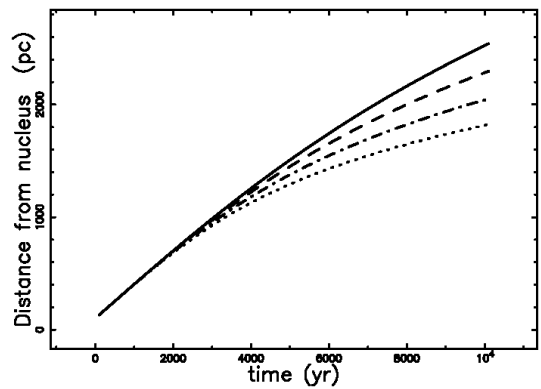}
\end {center}
\caption
{
Relativistic distance    as a function
of  time  when
$x_0$ =100~pc,
and
$\beta_0= 0.999$:
$\eta =0$     (full line),
$\eta =0.2$   (dashes),
$\eta =0.4$   (dot-dash-dot-dash)
and
$\eta =0.6$   (dotted).
}
\label{energyrelxt_abs}
    \end{figure*}

Figure \ref{energyrelxt_abs} 
shows the numerical trajectory
as a function of time for different values of the exponent $\eta$: 
an
increase in $\eta$ means a lower value for the traveled distance.

\subsection{The magnetic field}

The magnetic field  in CGS has an energy
density of $\frac{B^2}{8 \pi}$
where $B$ is the magnetic field.
The presence of the magnetic field can be modeled by adding
a second term for the density of energy
in the rest frame of the moving fluid, see
Eq.~(\ref{diffeqnrelpower})
which models the relativistic flow of energy  the  in presence
of an inverse power law
\begin{eqnarray}
({\rho_0\,{c}^{2}+\frac{B^2}{8\pi}) \left( {\frac {\rm d}{{\rm d}t}}x \left( t \right)
 \right) \pi \, \left( x \left( t \right)  \right) ^{2} \left( \tan
 \left( \frac{\alpha}{2} \right)  \right) ^{2} \left( {\frac {x_{{0}}}{x
 \left( t \right) }} \right) ^{\delta} \left( -{\frac { \left( {\frac
{\rm d}{{\rm d}t}}x \left( t \right)  \right) ^{2}}{{c}^{2}}}+1
 \right) ^{-1}} \nonumber \\
-({\rho_0\,{c}^{2}+\frac{B_0^2}{8\pi})   v_{{0}}\pi \,{x_{{0}}}^{2} \left( \tan
 \left( \frac{\alpha}{2} \right)  \right) ^{2} \left( -{\frac {{v_{{0}}}^{2}}{
{c}^{2}}}+1 \right) ^{-1}}=0
\label{eqndiffrelmag}
\quad  .
\end{eqnarray}

We continue assuming a constant of proportionality 
between the density of
energy of the magnetic field 
and the rest mass all along the jet
\begin{equation}
\frac{B(x)^2}{8 \pi} \propto \rho c^2 \propto (\frac{x_0}{x})^{\delta}
\quad .
\end{equation}
The magnetic field
as a function of the distance $x$ is
\begin{equation}
B =
\sqrt {{B_{{0}}}^{2} \left( {\frac {x_{{0}}}{x}} \right) ^{\delta}}
\label{bx}
\end{equation}
where $B_0$ is the magnetic field at $x=x_0$.
We  assume   an inverse power law  spectrum
for the ultrarelativistic  electrons
of the type
\begin{equation}
N(E)dE = K E^{-p} dE
\label{spectrum}
\end{equation}
where $K$ is a constant and $p$ the exponent of the inverse power law.
The intensity of the synchrotron radiation has a standard
expression, as given
by formula (1.175)  in \cite{lang2},
\begin{eqnarray}
I(\nu)
\approx 0.933 \times 10^{-23}
\alpha_p (p) K l  H_{\perp} ^{(p +1)/2 }
\bigl (
 \frac{6.26 \times 10^{18} }{\nu}
\bigr )^{(p-1)/2 }  \\
erg\, sec^{-1} cm^{-2} Hz^{-1} rad^{-2}
\nonumber
\end{eqnarray}
where $\nu$ is the frequency,
$H_{\perp}$ is the magnetic field perpendicular to the 
electron's velocity,
$l$ is the dimension of the radiating region
along the line of sight,
and  $ \alpha_p (p)$  is a slowly
varying function
of $p$ which is of the order of unity.
As an example, $p=2.5$ produces an intensity of the 
 type $I(\nu) \propto
\nu^{-0.75}$.

\begin{figure*}
\begin{center}
\includegraphics[width=7cm]{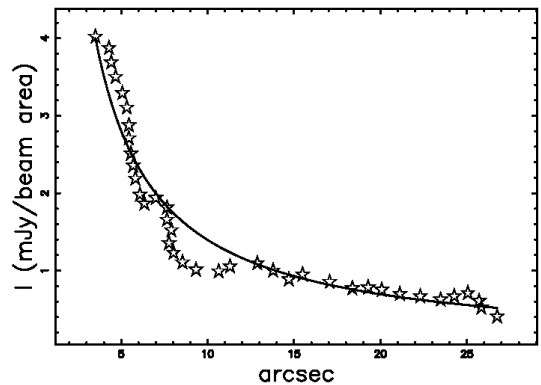}
\end {center}
\caption
{
Intensity profile along the centerline
of 3C31 when $x_0$ =3.51 arcsec,
$I_0=4$\ mJy/(beam\, area),
$p=2.5$,
$B_0= 10^{-4}$\ gauss,
$\delta=1.15$,
$\epsilon_{\mathrm {obs}} = 87.56\%$
and
$\chi^2$=3.05.
}
\label{intensity_3c31}
    \end{figure*}

We now analyse the intensity along the centerline of the jet,
which means
constant radiating length.
The intensity, assuming a constant $p$, scales as
\begin{equation}
I(x)={\frac {I_{{0}}{B(x)}^{p/2+1/2}}{{B_{{0}}}^{p/2+1/2}}}
\end{equation}
where $I_0$  is the intensity at  $x=x_0$ and  $B_0$ the magnetic field
at $x=x_0$.
We insert Eq.~(\ref{bx})  
in order to have an analytical
expression for the centerline intensity
\begin{equation}
I(x)=
{B_{{0}}}^{-p/2-1/2}I_{{0}} \left( {B_{{0}}}^{2} \left( {\frac {x_{{0}
}}{x}} \right) ^{\delta} \right) ^{p/4+1/4}
\label{intensitycenterline}
\end{equation}
and Figure \ref{intensity_3c31} 
shows the theoretical synchrotron
intensity as well the observed one in 3C31, see Figure 8 in
\cite{laing2002}.
We test the goodness of fit
through two standard  statistical tests.
The first test is the
$\chi^2$,
which is  computed as
\begin{equation}
\chi^2 =
\sum_{j=1}^n \bigl (I_{obs} -I_{theo}\bigr )^2
\label{chisquare}
\end{equation}
where
the  index $j$  varies  from 1 to the number of
available observations, $n$,
$I_{obs}$  is the  observed  intensity at position $j$,
and
$I_{theo}$ is
 the observed one.
A second  test of the  model works over
different points of the jet and
an
observational percentage  of
reliability,
 $\epsilon_{\mathrm {obs}}$,
is  introduced
\begin{equation}
\epsilon_{\mathrm {obs}}  =100\bigl (1-\frac{ \sum_j |I_{obs}-I_{theo}|_j }
                                      { \sum_j I_{theo,j}           }
                              \bigr )
.
\label{efficiencymany}
\end{equation}

\begin{figure*}
\begin{center}
\includegraphics[width=7cm]{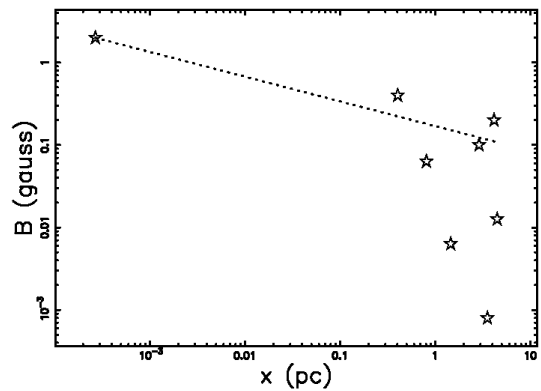}
\end {center}
\caption
{
Observed magnetic field density of 3C273 
as a function of the distance, empty stars,
and theoretical curve as represented by Eq.~(\ref{bx}), dotted line, 
when 
$ x_0 =2.6\times10^{-4}$ pc,
$ B_0 =2$ gauss,
$\delta=0.6$.
}
\label{bfield_3c273}
    \end{figure*}

Another application is to the spatial evolution
of the magnetic field of 3C273 as observed by VLBA in the pc region,
see \cite{Savolainen2008}.
Figure \ref{bfield_3c273}
shows the observed behaviour of the magnetic field 
as  well the theoretical evolution as represented by 
Eq.~(\ref{bx}).

\begin{figure*}
\begin{center}
\includegraphics[width=7cm]{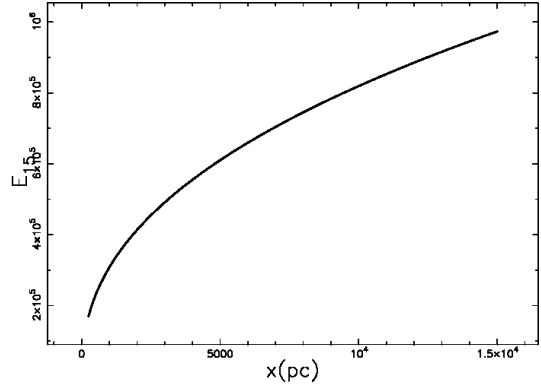}
\end {center}
\caption
{
Maximum achievable energy, $E_{15}$,
as a function of the distance
when
$x_0$ =100~pc,
$\beta_0= 0.9$,
$B_0= 0.025 $ gauss,
$\alpha=0.1$ and
$Z=1$.
}
\label{cosmicrays}
    \end{figure*}

The analytical expression for the magnetic field as 
a function of the
distance
allows finding the maximum energy  which can be reached in the
process of acceleration of the cosmic rays 
in extragalactic radio-sources.
The Hillas argument, see \cite{Hillas1984}, firstly introduces
the relativistic ions'  gyro-radius, $\rho_Z$, expressing
the energy  in $10^{15}$\ eV
units ($E_{15}$), 
the magnetic field in $10^{-6}$ gauss ($B_{-6}$)
\begin {equation}
\rho_Z = 1.08   \frac {E_{15}}  {B_{-6} Z  } pc
\label{rho}
\end   {equation}
where $Z$ is  the atomic number.
The relativistic gyro-radius is  
equalized to the maximum transversal
dimension of the jet, which is the diameter,
\begin{equation}
\rho_Z= 2 x  \tan ( \frac{\alpha}{2})
\quad .
\end{equation}
The resulting expression for the maximum energy is
\begin{eqnarray}
{E_{15}}=
 9.25\times 10^{5} x\tan \left(\frac{\alpha}{2} \right) \sqrt {{B_{{0
}}}^{2} \left( {\frac {x_{{0}}}{x}} \right) ^{\delta}}Z
\end{eqnarray}
where $B_0$ is expressed in gauss and $x$ and $x_0$ in pc.
Figure \ref{cosmicrays} 
reports the Hillas  plot for 3C31
from which it is possible to say that 
$E_{15}=10^6$ or $E=10^{21}$\ eV
can be reached at the end of the jet
when the magnetic field at $x_0 =100$\ pc
is  $B_0= 0.025$ gauss.

\section{Conclusions}

{\bf Classical turbulence:}
We modeled the physics of turbulent jets by the
conservation of the energy flux.
In the case of constant density,
we derived solutions for the  distance and velocity
as functions of
time, see Eqs~(\ref{energyxtconstant}) and
(\ref{energyvtconstant}).
In the presence of  an hyperbolic profile of density,
the solutions for the  distance and velocity
as functions of
time are  Eqs~(\ref{xthyperbolic}) and
(\ref{vthyperbolic}).
The case of a density which follows an inverse power
law of density
is limited to the derivation of the velocity, see
Eq.~(\ref{velocitypower}).
The presence of an inverse power law introduces
flexibility in the results and as an example
when $\delta=2$
the rate of mass flow
does not increases  with $x$ but is constant, see
Eq.~(\ref{mxpower}).

{\bf Relativistic turbulence:}
The conservation of the relativistic energy flux  for turbulent
jets is here analysed in  two cases.
In the first case we have a  surrounding medium 
with  constant density and 
the analytical result   is limited to
a series expansion for the solution, see Eq. (\ref{xtrelseries}).
In the second  case the surrounding density
decreases with a power law behaviour and   the analytical result
is limited to the velocity--distance relation,
see Eq. (\ref{betadistance})
and to a series expansion for the solution, 
see Eq. (\ref{xtseriesreldelta}).

{\bf The losses:}
The choice of the flux of energy as a quantity to be conserved
allows a parametrization of the losses. In the first
model
we considered
the decrease of the available classical and relativistic
flux of energy through a recursive relation, see
Eqs~(\ref{eqnrecursive})   and  (\ref{eqnrecursiverelativistic}).
Figures  \ref{recursive_classic}
and      \ref{recursive_rel}
show
the velocity as a function of the regulating parameter $\varepsilon$.
Values of  $\varepsilon<0.001$ do  not affect the jet's trajectory
at the astrophysical distance of 15 kpc.
In the second model, we fixed a law for the decrease of the available
flux of relativistic energy as a function of the distance,
see Eq.~(\ref{eqndiffrellosses})
 and we
derived a law for the decrease of the velocity  as a function of the
regulating parameter $\eta$, see Figure \ref{energyrelxt_abs}.

{\bf  Astrophysical applications:}
We modeled the behaviour    of the magnetic field
assuming the  conservation of the magnetic flux of energy in the
case of constant density,
see  Eq.~(\ref{eqndiffrelmag}).
The availability of an
analytical expression for the magnetic field,  see the theoretical 
Eq.~(\ref{bx}),
allows findinging a law for the behaviour of the intensity
of the synchrotron emission, see Eq.~(\ref{intensitycenterline}).
The application to the measured intensity of  3C31
yields an  efficiency over all the jet's length  of $87.56\%$,
see  Figure \ref{intensity_3c31}.
A  test  on the magnetic field of
3C273 in the pc region can be seen in Figure \ref{bfield_3c273}. 
The presence of a law  for the magnetic field
allows fixing the Hillas plot for the maximum energy
which can reached during the process of acceleration
of the cosmic rays, which in the case of 3C31
is $\approx 10^{21}$\ eV, see the caption of Figure \ref{cosmicrays}.

{\bf References}

\providecommand{\newblock}{}

\providecommand{\newblock}{}
\end{document}